\documentclass[10pt,journal]{IEEEtran}

\usepackage[dvips]{graphicx}
\usepackage[cmex10]{amsmath}

\renewcommand{\baselinestretch}{1.16}

\title{Joint DOA and Array Manifold Estimation for a MIMO Array Using Two Calibrated Antennas}

\author{Wei Zhang{\small $~^{\#}~^{*}$}, Wei Liu{\small $~^{*}$}, Siliang Wu{\small $~^{\#}$}, and Ju Wang{\small $~^{\#}$}\\ \vspace*{0.4cm}
        $~^{\#}${\normalsize Department of Information and Electronics}\\
        {\normalsize Beijing Institute of Technology, Beijing, China}\\\vspace*{0.4cm}
       $~^{*}$   {\normalsize Communications Research Group}\\ {\normalsize Dept. of Electronic \& Electrical Engineering}\\
        {\normalsize University of Sheffield, UK}}
        
\begin{document}

\maketitle
\begin{abstract}
  A simple scheme for joint direction of arrival (DOA) and array manifold estimation for a MIMO array system is proposed, where only two transmit antennas are calibrated initially. It first obtains a set of initial DOA results by employing a rotational invariance property between two sets of received data, and then more accurate DOA and array manifold estimation is obtained through a local searching algorithm with several iterations. No strict half wavelength spacing is required for the uncalibrated antennas to avoid the spatial aliasing problem.
\end{abstract}


\section{Introduction}

A MIMO radar array system employs multiple transmit antennas for emitting orthogonal waveforms and multiple receive antennas for receiving the echoes reflected by the targets \cite{Li2009,Fishler2006,Hassanien2010} and can exploit the waveform diversity to form a virtual array with increased degrees of freedom (DOFs) and a larger aperture compared to the traditional phased-array radar. It has been shown that MIMO radar can provide enhanced spatial resolution, achieve better target detection performance, and significantly improve the system's parameter identifiability \cite{Hassanien2010,ZhangWei2013,Li2007,liu14a}.

Many techniques have been proposed for angle estimation in MIMO radar using traditional direction of arrival (DOA) algorithms, such as MUSIC \cite{Schmidt1986} and ESPRIT~\cite{Roy1989},  by assuming perfect knowledge of the array manifold. However, these algorithms are sensitive to uncertainties in the array manifold, and their performance will degrade significantly in the presence of array model errors~\cite{Friedlander1990,Parvazi2011,See2004}. On the other hand, it is time-consuming and expensive to calibrate the system in the case of large or time-varying arrays \cite{Parvazi2011}. In addition, it is observed that in practice, even after initial calibration, antenna gain and phase errors still exist due to environmental changes and other factors \cite{Soon1994}. To cope with the problem, in \cite{Friedlander1991}, a MUSIC-based DOA estimation method in the presence of gain and phase errors was introduced. A subspace-based method for estimating the errors was proposed in \cite{Soon1994}. Other methods were also proposed based on partially calibrated arrays \cite{Weiss1996,See2004,Parvazi2011}.
Additionally, blind calibration is possible for non-Gaussian signals
by using higher-order statistics \cite{Kim2010}, although with a very high computational complexity.

In this work, we address the problem of joint DOA and array manifold estimation with a multi-input multi-output (MIMO)  array configuration~\cite{Fishler2006,Li2007},
where only two transmit antennas are fully calibrated, while the receive antennas are uncalibrated~\cite{liu13a,liu13c}. Since the two transmit antennas transmit orthogonal waveforms, we can extract the received data associated with each transmit antenna.
With the two transmit antennas well calibrated, a rotational invariance property between the two sets of data can still be maintained without any knowledge of the array manifold of the uncalibrated side; then the ESPRIT algorithm can be used to find the initial DOAs of the targets. Starting with the initial DOA estimates, the antenna gains and phases can then be estimated through an appropriate modification of the MUSIC algorithm introduced in \cite{Friedlander1991}. The estimated antenna gains and phases will be used in the more accurate estimation of DOAs via the MUSIC algorithm. This procedure will be repeated until some convergence criterion is met.
The advantage of the scheme is that only two calibrated antennas are needed for high resolution DOA estimation and no specific requirement is imposed on the uncalibrated antennas. To our best knowledge, none of the existing DOA estimation methods  for MIMO arrays has considered the joint DOA and array manifold estimation problem.

This paper is organized as follows. In Sec. II, the array model and a review of DOA estimation are provided, with the proposed method given in Sec. III. Simulation results are presented in Sec. IV and conclusions are drawn in Sec. V.

%%%%%%%%%%%%%%%%%%%%%%%%%%%%%%%%%%%%%%%%%%%%%
\section{Background}
%\subsection{Array Model}
   Consider a MIMO system with a uniform linear array (ULA) of $M$ antennas used for both transmitting and receiving.
 For simplicity of notation and without loss of generality, we assume that the first two antennas are perfectly calibrated.
 The steering vector of the ULA is then given by
   \begin{eqnarray}
   \textbf{a}(\theta)=[1,e^{-j2\pi d\sin(\theta)/\lambda},\alpha_3e^{j\phi_3}e^{-j2\pi 2d\sin(\theta)/\lambda}, \cdots,  \nonumber \\
   \alpha_Ne^{j\phi_N} e^{-j2\pi(M-1) d\sin(\theta)/\lambda}]^T
   \label{eq3}
   \end{eqnarray}
    where $[\cdot]^T$ denotes the transpose operation, $\theta$ is the angle of the pointing direction, $d$
      is the inter-element spacing, $\lambda$ is the signal wavelength, and $\alpha_i$ and $\phi_i$ denote the gain and phase errors, respectively.
   Assume that $K$ targets are present. The output of the
   matched filters at the receiver is given by \cite{Li2007}
\begin{eqnarray}
\textbf{x}[n]&=&\sum_{k=1}^{K}\textbf{a}(\theta_k)\otimes \textbf{a}(\theta_k)b_k[n]+\textbf{n}[n] \nonumber\\
%&=&[\textbf{a}(\theta_1)\otimes \textbf{a}(\theta_1),\textbf{a}(\theta_2)\otimes \textbf{a}(\theta_2),\cdots, \nonumber \\
%&~& \textbf{a}(\theta_K)\otimes \textbf{a}(\theta_K)]\textbf{b}[n]+\textbf{n}[n]\nonumber\\
&=&\textbf{A}\textbf{b}[n]+\textbf{n}[n]
\label{eq1}
\end{eqnarray}
where $\theta_k$ is the DOA of the $k$th target, $\otimes$ is the Kronecker product,
$b_k[n]=\beta_ke^{j2\pi f_dn}$, with  $\beta_k$ being the complex-valued reflection coefficient of the $k$th target and $f_d$ being the Doppler frequency, $\textbf{b}[n]=[b_1[n],b_2[n],\cdots,b_K[n]]^T$,
\begin{equation}
\textbf{A}=[\textbf{a}(\theta_1)\otimes \textbf{a}(\theta_1),\cdots,
\textbf{a}(\theta_K)\otimes \textbf{a}(\theta_K)]
\label{eq4}
\end{equation}
  is the overall transmit-receive or virtual array manifold, and
 $\textbf{n}[n]$ is the white noise vector with a power $\sigma^2$.

Assume that all target-reflected signals and noise are uncorrelated. Then we have
     \begin{eqnarray}
     \textbf{R}_{x}&=&E[\textbf{x}[n]\textbf{x}[n]^H]=\textbf{A}\textbf{R}_b\textbf{A}^H+\sigma^2\textbf{I} \nonumber \\
     &=&\textbf{U}_s\varLambda\textbf{U}_{s}^H+\sigma^2
             \textbf{U}_n\textbf{U}_{n}^H
     \label{eq5}
     \end{eqnarray}
where $E[\cdot]$ and $[\cdot]^H$ denote expectation  and Hermitian transpose, respectively, $\textbf{R}_b=E[\textbf{b}[n]\textbf{b}[n]^H]$, $\varLambda=\mathrm{diag}\{\lambda_1,\cdots,\lambda_K\}$ consists of the $K$ principal eigenvalues of  $\textbf{R}_{x}$,
       $\textbf{U}_s$ is the signal subspace, specified by the principal eigenvectors of $\textbf{R}_{x}$, and the remaining eigenvectors $\textbf{U}_n$ is the noise subspace.
  In practice, $\textbf{R}_{x}$ will be replaced by
$\hat{\textbf{R}}_{x}=\frac{1}{L}\sum_{n=1}^{L}\textbf{x}[n]\textbf{x}[n]^H$,
  where $L$ is the number of snapshots.

% \subsection{DOA estimation for MIMO radar}
  The MUSIC algorithm for DOA estimation for MIMO radar can be constructed as \cite{Zhang2010a,He2011}
  \begin{equation}
  f(\theta)=1/[\textbf{a}(\theta)\otimes \textbf{a}(\theta)]^H\textbf{U}_n\textbf{U}_{n}^H [\textbf{a}(\theta)\otimes \textbf{a}(\theta)].
  \label{eq6a}
  \end{equation}
    The $K$ largest peaks of $f(\theta)$ indicate the DOAs of the targets. It requires the spacing between two adjacent
    antennas to be within a half wavelength to avoid estimation
    ambiguity.

   For ESPRIT estimator \cite{Duofang2008}, it is based on the signal subspace $\textbf{U}_s$. Let $\textbf{U}_{s,1}$ be the subset of $\textbf{U}_s$, which relates to the first to the $(M-1)$-th transmit antennas, and $\textbf{U}_{s,2}$ be the subset of $\textbf{U}_s$, which relates to the second to the $M$-th transmit antennas.
   We then have the following relationship
    \begin{equation}
   \textbf{U}_{s,2}=\textbf{U}_{s,1}\textbf{T}_e\textbf{Q}_e\textbf{T}^{-1}_e
   \end{equation}
    where $\textbf{T}_e$ is an unknown nonsingular matrix and $\textbf{Q}_e$ is a diagonal matrix, with its $k$th main diagonal element being $e^{-j2\pi d\sin(\theta_k)/\lambda}$.
   Thus, the DOAs can be found from the eigenvalues of $(\textbf{U}_{s,1}^H\textbf{U}_{s,1})^{-1}\textbf{U}_{s,1}^H\textbf{U}_{s,2}$.

%%%%%%%%%%%%%%%%%%%%%%%%%%%%%%%%%%%%
  \section{Proposed Method}
  In this section, we first perform an initial DOA estimation using the two sets of received data associated with the first and the second transmit antennas by applying the ESPRIT algorithm, then the gain and phase errors can be estimated using the initial DOA results by applying a MUSIC-based approach.
  \subsection{Estimating initial DOAs}
  Since the array manifold  is unknown, we can not apply the traditional subspace-based methods directly. To solve the problem, define $\textbf{A}_1$ and $\textbf{A}_2$ as the first and the second $M$ rows of $\textbf{A}$, respectively, with
   \begin{eqnarray}
   \textbf{A}_1&=&[\textbf{a}(\theta_1),\cdots,\textbf{a}(\theta_K)],  \\
   \textbf{A}_2&=&[e^{-j2\pi d\sin(\theta_1)/\lambda}\textbf{a}(\theta_1), \cdots, e^{-j2\pi d\sin(\theta_K)/\lambda}\textbf{a}(\theta_K)] \nonumber \\
   &=&\textbf{A}_1\textbf{Q}
   \end{eqnarray}
  where $\textbf{Q}$ is an $M\times M$ diagonal matrix, with $e^{-j2\pi d\sin(\theta_k)/\lambda}$ being its $k$th main diagonal element.

Although there are model errors in both $\textbf{A}_1$ and $\textbf{A}_2$,
a rotational invariance property between $\textbf{A}_1$ and $\textbf{A}_2$ is still maintained, which enables the use of ESPRIT for DOA estimation.
%Let $\textbf{U}_s$ be the signal subspace composed of the principal eigenvectors corresponding
%to the $K$ largest eigenvalues of $\hat{\textbf{R}}_{x}$.
$\textbf{A}$ and $\textbf{U}_s$ have a relationship determined by a unique nonsingular matrix $\textbf{T}$ as
\begin{equation}
\textbf{A}=\textbf{U}_s\textbf{T}.
\end{equation}

   Define $\textbf{U}_1$ and $\textbf{U}_2$ as the first and second $M$ rows of $\textbf{U}_s$, respectively. We have
   \begin{eqnarray}
   \textbf{A}_1&=&\textbf{U}_1\textbf{T}, \label{eq11}\\
   \textbf{A}_2&=&\textbf{U}_2\textbf{T}=\textbf{A}_1\textbf{Q}.
   \label{eq12a}
   \end{eqnarray}
  Then,
   \begin{equation}
   \textbf{U}_2=\textbf{U}_1\textbf{T}\textbf{Q}\textbf{T}^{-1}.
   \end{equation}
   Now using the traditional ESPRIT technique, the main diagonal elements of $\textbf{Q}$ can be obtained via eigendecomposition of $(\textbf{U}_1^H\textbf{U}_1)^{-1}\textbf{U}_1^H\textbf{U}_2$. Since the two transmit antennas have been well calibrated, $\{\theta_k\}_{k=1}^{K}$ can be obtained easily from $\textbf{Q}$.

   Note that  the rotational invariance property exploited here depends only on the two calibrated transmit antennas and is not related  to the uncalibrated part. Thus, the initial DOAs can be estimated accurately without any
   knowledge of array model errors. Additionally, in this initial DOA estimation, the proposed ESPRIT-based method imposes less constraints on the spacing of the uncalibrated part, which can be arranged to be much larger than a half-wavelength for a high-resolution DOA estimation.

   \subsection{Estimating array manifold}
   From (\ref{eq6a}), with exactly known $\textbf{R}_x$, the DOAs can also be found by solving the following equation \cite{Friedlander1991}:
     \begin{equation}
    [\textbf{a}(\theta)\otimes\textbf{a}(\theta)]^H\textbf{U}_n\textbf{U}_{n}^H[\textbf{a}(\theta)\otimes\textbf{a}(\theta)]=0.
      \label{eq10}
     \end{equation}
      The actual steering vector can also be expressed as
               \begin{equation}
                  \textbf{a}(\theta)=\Gamma \bar{\textbf{a}}(\theta)
                  \label{eq12}
                  \end{equation}
                  where $\Gamma=\mathrm{diag}[1,1,\alpha_3e^{j\phi_3},\cdots,\alpha_Me^{j\phi_M}]$ and $\bar{\textbf{a}}(\theta)=[1,e^{-j2\pi d\sin(\theta)/\lambda}, \cdots,
                        e^{-j2\pi(M-1)d\sin(\theta)/\lambda}]^T$.
    Therefore, the estimate of antenna gains and phases can be obtained using the initially estimated DOAs as follows:
       \begin{eqnarray}
       &&\min\sum_{k=1}^{K}\big[\big(\Gamma\bar{\textbf{a}}(\hat{\theta}_k)\big)\otimes\big(\Gamma\bar{\textbf{a}}(\hat{\theta}_k)\big)\big]^H\textbf{U}_n\textbf{U}_{n}^H\big[\big(\Gamma\bar{\textbf{a}}(\hat{\theta}_k)\big)\nonumber \\
      && ~~~~~\otimes\big(\Gamma\bar{\textbf{a}}(\hat{\theta}_k)\big)\big]  \nonumber \\
      && =\min_{\delta}\sum_{k=1}^{K}[ \textbf{V}_k\delta]^H\textbf{U}_n\textbf{U}_{n}^H[\textbf{V}_k\delta]  \nonumber \\
    && ~ ~~ \mathrm{subject~to}~\delta^H\textbf{e}_1=1,~\delta^H\textbf{e}_2=1
    \label{eq13}
       \end{eqnarray}
       where $\delta$ is the $M^2\times 1$ gain and phase vector, with its elements being the diagonal elements of $[\Gamma\otimes\Gamma]$, $\textbf{V}_k=\mathrm{diag}[\bar{\textbf{a}}(\hat{\theta}_k)\otimes\bar{\textbf{a}}(\hat{\theta}_k)]$, with $\hat{\theta}_k$ being the initial DOA estimate of the $k$th target, $\textbf{e}_1=[1,0,\cdots,0]^T$ and $\textbf{e}_2=[0,1,0,\cdots,0]^T$. It should be noted that both the $(M+1)$-th and the $(M+2)$-th elements of $\delta$ should also be equal to 1; however, we find that the above two constraints are able to give a satisfactory result.
  The problem in (\ref{eq13}) can be rewritten as
  \begin{eqnarray}
        && \min_{\delta}\delta^H\textbf{Z}\delta  \;\;\; \mathrm{subject~to}~\delta^H\textbf{e}=f^T
      \label{eq14}
         \end{eqnarray}
  where $\textbf{Z}=\sum_{k=1}^{K}\textbf{V}_{k}^H\textbf{U}_{n}\textbf{U}_{n}^H\textbf{V}_{k}$, $\textbf{e}=[\textbf{e}_1,\textbf{e}_2]$, and $f=[1,1]^T$. Its solution is given by
  \begin{equation}
  \delta=\textbf{Z}^{-1}\textbf{e}[\textbf{e}^H\textbf{Z}^{-1}\textbf{e}]^{-1}f^T.
  \label{eq15}
  \end{equation}

 Using the estimates (\ref{eq15}), the DOAs can be estimated from the $K$ highest peaks of the following function:
   \begin{equation}
     f(\theta)=\frac{1}{\big[\mathrm{diag}[\delta][\bar{\textbf{a}}(\theta)\otimes\bar{\textbf{a}}(\theta)]\big]^H\textbf{U}_n\textbf{U}_{n}^H\big[\mathrm{diag}[\delta][\bar{\textbf{a}}(\theta)\otimes\bar{\textbf{a}}(\theta)]\big]}
      \label{eq16}.
  \end{equation}
 Since a set of initial DOA estimates has already been obtained, we can search for each DOA estimate over a small DOA region corresponding to each initial DOA estimate. Thus, the inter-element spacing of the uncalibrated  array does not have to be smaller than half wavelength to avoid estimation ambiguity. Actually, we can increase the inter-element spacing of the uncalibrated array to improve the accuracy of estimation.

 The proposed joint DOA and array manifold estimation scheme is summarized as follows:

 1) Estimate the initial DOAs using the ESPRIT algorithm.

 2) Estimate the array manifold using (\ref{eq15}).

 3) Use the results in Step 2 to find updated DOAs by local searching through  (\ref{eq16}).

 4) Repeat Steps 2 and 3 until some convergence criterion   is satisfied. One such a  criterion could be the difference between the estimation results of the last round and the current one. When this difference is smaller than a pre-set threshold value, we can then stop the iteration.

Note that we have assumed implicitly that the antenna positions have been calibrated, and we consider the fixed uncalibrated gain and phase errors only. This is because the calibration of array position is more convenient than the calibration of gain and phase which may vary due to environmental changes. On the other hand, the position error can be transformed into phase errors. However, the phase errors caused by position errors are not fixed for the targets because the targets have different DOAs. In such a case, a simple way is to obtain the gain and phase errors corresponding to each target, i.e. we should estimate the gain and phase errors when obtaining one target's DOA other than all the DOAs.

\subsection{Complexity analysis}
To estimate the sample covariance matrix, a computational complexity of $O(M^4L)$ is needed.
The eigendecomposition operation needs a computational complexity of $O(M^6)$.
The proposed ESPRIT requires a computational complexity of $O(M^3)$.
In the estimation of array manifold, the computational complexity of $O(M^6n)$ is needed, where $n$ is the iteration number.
 Therefore, the proposed scheme has at least a complexity of $O(M^6n+M^6+M^4L+M^3)$.
\subsection{Cram\'{e}r-Rao Bound for Uncalibrated Array}
  In this section, we derive the stochastic CRB for uncalibrated array by extending the results of \cite{See2004,Nehorai1994a}.
Define
%   \begin{equation}
  $ h_{i}=\alpha_ie^{j\phi_i}$, i=3, $\cdots$, M,
%   \end{equation}
   as the gain and phase error that corresponds to the $i$th sensor and the $(2M-4+K)\times 1$ vector $\pmb{\eta}=[\pmb{\theta}^T,\pmb{\xi}^T,\pmb{\zeta}^T]^T$ containing the unknown parameters,
                where
                \begin{eqnarray}
                \pmb{\theta}&=&[\theta_1,\cdots,\theta_K]^T\\
               \pmb{\xi}&=&[\mathrm{Re}\{h_{3} \},\cdots,\mathrm{Re}\{h_{M}  \}]^T \\
               \pmb{\zeta}&=&[\mathrm{Im}\{h_{3}  \},\cdots,\mathrm{Im}\{ h_{M} \}]^T.
                \end{eqnarray}
The snapshots are assumed to satisfy the stochastic model
   \begin{equation}
   \textbf{x}[n]=\mathcal{N}\{\textbf{0},\textbf{R}_x \}
   \end{equation}
   where $\mathcal{N}\{\cdot,\cdot\}$ is the complex Gaussian distribution.
   The unknown parameters include the elements of $\pmb{\eta}$, the noise variance $\sigma^2$, and the parameters of the source covariance
   matrix $\{[\textbf{R}_b]_{ii}\}_{i=1}^{K}$ and $\{\mathrm{Re}\{[\textbf{R}_b]_{ij}\}, \mathrm{Im}\{[\textbf{R}_b]_{ij}\};j>i\}_{i,j=1}^K$.

    Considering the problem with respect to the parameters of
     the source covariance matrix and the noise variance, the $(2M-4+K)\times(2M-4+K)$ Fisher information matrix can be written as \cite{See2004,Nehorai1994a}
     \begin{equation}
     [\textbf{F}(\pmb{\eta})]_{i,j}=
     \frac{2L}{\sigma^2}\mathrm{Re}
     \Big\{\mathrm{trace}\Big(\textbf{W}\frac{\partial \textbf{A}^H}{\partial \eta_j}\textbf{P}_{\textbf{A}}^{\bot}\frac{\partial \textbf{A}}{\partial \eta_i}           \Big) \Big\}
     \label{eq17}
     \end{equation}
     where $\textbf{P}_{\textbf{A}}^{\bot}=\textbf{I}-\textbf{A}(\textbf{A}^H\textbf{A})^{-1}\textbf{A}^H$ is the $M\times M$ orthogonal projection matrix and the $K\times K$ matrix
       $
       \textbf{W}=\textbf{R}_b(\textbf{A}^H\textbf{A}\textbf{R}_b+\sigma^2\textbf{I})^{-1}\textbf{A}^H\textbf{A}\textbf{R}_b
       $.
       Then the CRB matrix is $\textbf{CRB}=\textbf{F}^{-1}$.

%%%%%%%%%%%%%%%%%%%%%%%%%%%%%%%%%%%%%
\section{Simulations}
Simulations are carried out to investigate the performance of the proposed method compared with the traditional ESPRIT estimator in \cite{Duofang2008} and the MUSIC estimator. We consider a MIMO array with $M=10$ antennas and half-wavelength spacing. The first two antennas are perfectly calibrated. $K=3$ targets are located at $10^\circ$, $20^\circ$, and $30^\circ$, respectively.
Results from $100$ simulation runs are averaged to give the root mean square error (RMSE) of the estimates. %assess the overall
For all simulations, the number of snapshots $L=100$ is used.
%\subsection{Example 1}
\begin{figure}[htbp]
\small
\centering
\includegraphics[angle=0,width=0.48\textwidth]{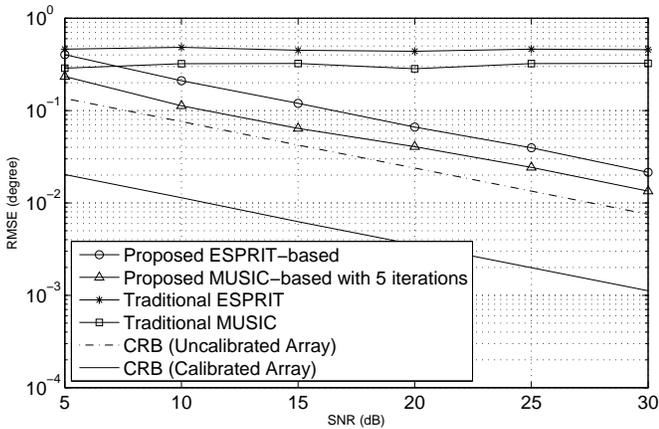}
\caption{RMSEs of DOA estimation versus input SNR.}
\label{fig1}
\end{figure}
We first study the performance of the proposed ESPRIT-based algorithm for initial DOA estimation. The antenna gain and phase errors are assumed to have a uniform distribution: $\alpha_{k}\in[0.8,1.2]$ and $\phi_{k}\in[-\pi/10,\pi/10]$.  $\alpha_{k}$ and $\phi_{k}$ change from run to run while remaining constant for all snapshots.
Fig. \ref{fig1} shows the RMSE results versus input SNR. We see that the gain and phase errors have significantly degraded the performance of the traditional MUSIC and ESPRIT algorithms. However, the proposed one is quite robust and has a much better performance. In this figure, we also showed the result of our proposed method with $5$ iterations, and a clear improvement can be observed compared to the initial estimation.

\begin{figure}[htbp]
\small
\centering
\includegraphics[angle=0,width=0.48\textwidth]{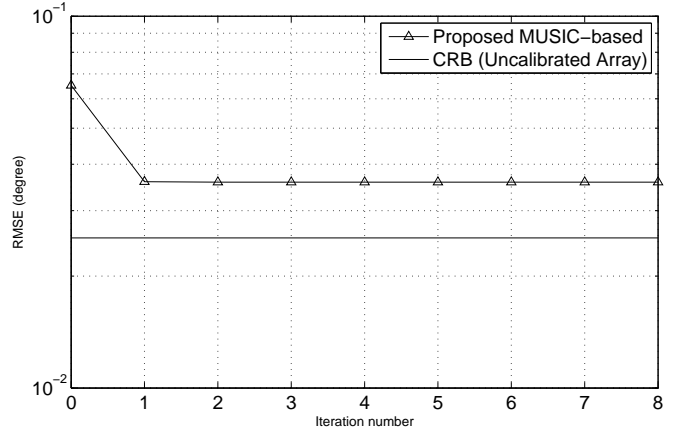}
\caption{RMSEs of DOA estimation versus iteration number.}
\label{fig2}
\end{figure}

\begin{figure}[htbp]
\small
\centering
\includegraphics[angle=0,width=0.48\textwidth]{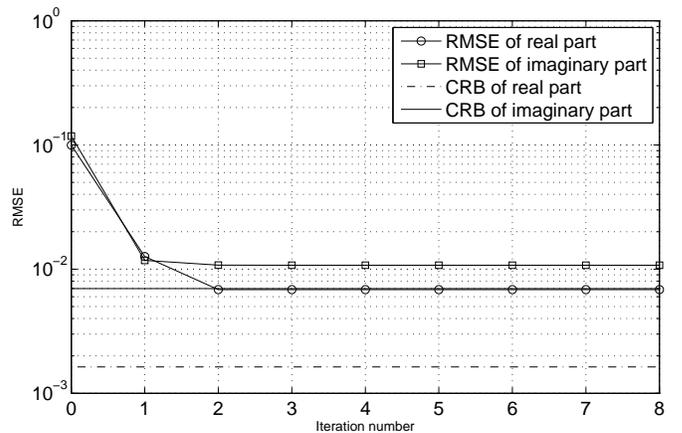}
\caption{RMSEs of gain and phase estimation versus iteration number.}
\label{fig3}
\end{figure}

In the second example, the effect of the iteration number on the performance of the proposed method is demonstrated. The input SNR is set to 20 dB and the antenna gain and phase errors are set as (the diagonal elements of $\Gamma$)

\begin{eqnarray}
 [1,1,1.13e^{-j0.020},0.89e^{j0.180},1.1e^{j0.130},1.05e^{-j0.038},\nonumber\\
 0.98e^{j0.101},0.90e^{-j0.057},1.15e^{-j0.187},0.88e^{-j0.247}].
\end{eqnarray}
The RMSE for DOA estimation versus the iteration number is shown in Fig. \ref{fig2} and the result for unknown parameters estimation is shown in Fig. \ref{fig3}. Clearly the first or two iterations have already led to an accurate enough result.

\begin{figure}[htbp]
\small
\centering
\includegraphics[angle=0,width=0.48\textwidth]{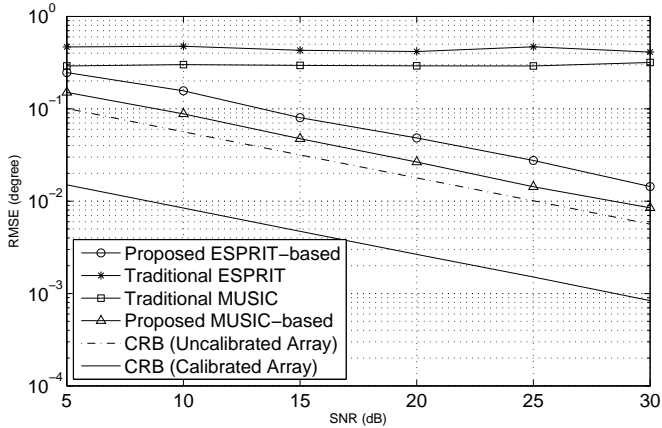}
\caption{RMSEs of DOA estimation versus input SNR.}
\label{fig5}
\end{figure}

Now we study the effect of antenna spacing on the performance of the proposed method with $5$ iterations.
 The spacing between the two calibrated antennas is $0.5\lambda$, while the spacing between the uncalibrated antennas is set to $2\lambda$ for the proposed method,
 and $0.5\lambda$ for the other methods. The other parameters remain the same as in Example 1.
The results are shown in Fig. \ref{fig5}. We can see that the proposed ESPRIT-based initial estimation has achieved a higher accuracy compared to Fig. \ref{fig1}, and the performance of the proposed method is much better than the corresponding result of Example 1 and significantly outperforms the other considered algorithms.

\section{Conclusions}
A joint DOA and array manifold estimation scheme for a MIMO array system has been proposed, where only two antennas at the transmit side are initially calibrated, while the remaining part of the system is completely uncalibrated. By exploiting the rotational invariance property between two sets of received data associated with the two calibrated antennas, the ESPRIT algorithm is first employed to give a set of initial DOA estimation results, which is then used by the following MUSIC-based algorithm for the joint estimation.
Additionally, the proposed scheme does not require the adjacent antenna spacing in the uncalibrated part to be within a half wavelength, which provides  further improvement to the estimation.

\end{document}